\documentclass[preprintnumbers, floatfix,letterpaper, twocolumn,aps,prd,epsfig]{revtex4-1}
\usepackage{bm,graphicx,dcolumn,epstopdf,epsf, latexsym,mathbbol, amssymb,amsmath,color,slashed, mathrsfs,mathcomp,simplewick}
\usepackage[center]{subfigure}
\usepackage{multirow}
\usepackage{makecell}
\usepackage[colorlinks,linkcolor=blue,citecolor=blue,urlcolor=blue]{hyperref}
\usepackage[center]{subfigure}

\begin{document}
\allowdisplaybreaks
 \newcommand{\bq}{\begin{equation}}
 \newcommand{\eq}{\end{equation}}
 \newcommand{\bqn}{\begin{eqnarray}}
 \newcommand{\eqn}{\end{eqnarray}}
 \newcommand{\ban}{\begin{align}}
 \newcommand{\ean}{\end{align}}
  \newcommand{\nb}{\nonumber}
 \newcommand{\lb}{\label}
 \newcommand{\f}{\frac}
 \newcommand{\p}{\partial}
\newcommand{\PRL}{Phys. Rev. Lett.}
\newcommand{\PLB}{Phys. Lett. B}
\newcommand{\PRD}{Phys. Rev. D}
\newcommand{\CQG}{Class. Quantum Grav.}
\newcommand{\JCAP}{J. Cosmol. Astropart. Phys.}
\newcommand{\JHEP}{J. High. Energy. Phys.}
\newcommand{\NPB}{Nucl. Phys. B}
\newcommand{\Doi}{https://doi.org}

\title{Accretion disk around Reissner-Nordstr\"{o}m black hole coupled with a nonlinear electrodynamics field}

\author{ G. Abbas ${}^{a,b}$}
\email{ghulamabbas@iub.edu.pk}

\author{H. Rehman  ${}^{a}$}
\email{hamzarehman244@gmail.com}

\author{Tao Zhu ${}^{c, d}$}
\email{zhut05@zjut.edu.cn}

\author{Qiang Wu ${}^{c, d}$}
\email{wuq@zjut.edu.cn}

\author{G. Mustafa ${}^{e, f}$}
\email{gmustafa3828@gmail.com}

\affiliation{
${}^a$ National Astronomical Observatories, Chinese Academy of Sciences, Beijing 100101, China\\
${}^{b}$ Department of Mathematics, The Islamia University of Bahawalpur, Bahawalpur Pakistan \\
${}^{c}$ Institute for Theoretical Physics and Cosmology, Zhejiang University of Technology, Hangzhou, 310032, China\\
${}^{d}$ United Center for Gravitational Wave Physics (UCGWP), Zhejiang University of Technology, Hangzhou, 310032, China
${}^{e}$ United Center for Gravitational Wave Physics (UCGWP),  Zhejiang University of Technology, Hangzhou, 310023, China\\
${}^{f}$ New Uzbekistan University, Mustaqillik Ave. 54, Tashkent 100007, Uzbekistan}


\date{\today}

\begin{abstract}

The phenomenon by which matter accumulates in the vicinity of a huge celestial object is known as accretion. The gravitational energy is excreted as a consequence of infalling matter onto compact objects. The accretion procedure around celestial bodies like neutron stars, white dwarfs, and black holes has considerable importance because of its ability to transform gravitational energy into radiation. This study investigates the particle's geodesic motion and accretion around the spherically symmetric Reissner-Nordstr\"{o}m black hole coupled with a nonlinear electrodynamics field utilizing isothermal fluid. The formation of the disc-like structure in the accretion process arises from the geodesic motion exhibited by particles near the black hole. The circular orbits, radiant flux energy, radioactive efficiency, and radiant temperature, can be determined. Our study focuses on the examination of particles exhibiting stable circular orbits within the equatorial plane. We analyze the perturbations experienced by particles throughout employing restoring forces and the oscillatory behavior of the particles around a compact object. We conduct an analysis of the fluid's critical flow and maximum accretion rate. Our results show how the black hole parameter $\zeta$ and charge $q$ affect the circular geodesic of particles and the maximum accretion rate of the Reissner-Nordstr\"{o}m black hole coupled with nonlinear electrodynamics.

\end{abstract}


\maketitle

\section{Introduction}

The theory of general relativity (GR) speculates on the presence of black holes (BHs) as mysterious objects. The extremely strong gravitational field in the universe is considered to have originated from BH. Additionally, it is believed that BHs possess strong magnetic fields and spin. In light of these characteristics, BHs are the ideal astrophysical laboratory for studying the nature of gravity and the matter around it. Based on the examination of observational evidence, empirical data has recently confirmed the existence of BH. The first accomplishment represented the discovery of gravitational waves arising due to the collision of two BHs in a  binary system, as observed through the collaborative work of LIGO and Virgo \cite{a1}.  Another significant role of the Event Horizon Telescope is its utilization of baseline Interferometry to capture the first images of the BH shadow of M87 \cite{a2, a3}, as well as revealed image of Sgr A$^{*}$ \cite{a4}.

It is believed that cosmic entities, such as BHs, undergo mass accumulation via the phenomenon of accretion. They might also serve to analyze modified theories of gravity. The existence of an accretion disc is an essential element in pursuing the higher accretion rate encompassing these compact objects. Diffuse matter creates the accretion disc and emits energy by slowly spiraling into a centrally condensed object. Accretion is the process by which a fluid nearby attracts particles to a compact object like a BH. Whenever the fluid velocity is identical to the sound speed, these particles must pass through the critical point. The fluid is projected onto its central mass at supersonic speeds. The BH mass needs to be raised as a result of this event. \cite{a5}. It is fascinating to analyze numerous usual radii as a consequence of examining the particles' geodesic structure near the BH, such as the innermost stable circular orbit (ISCO) and marginally bound orbit ($r_{mb}$). In the examination of BH accretion discs, the considered radii are the significant factors.

The ISCO is associated with the inner boundary of the accretion disc around a BH, and their radii can be used to compute the energy emission efficiency, which is a measure of how quickly energy from the rest mass turns into radiation. The locations of unstable or stable circular orbits correspond to the greatest or lowest value of the effective potential, accordingly. According to Newtonian theory, it is believed that the ISCO does not have a minimum radius. This is supported by the observation that the ISCO can assume any radius once the effective potential reaches its smallest value for all possible values of angular momentum \cite{a8}. For any minimum or maximum value of the angular momentum, the effective potential in GR and particles rotating near the Schwarzschild BH comprises two extremes, so it corresponds with the two points. One can explored ISCO at $r = 3r_{g}$ \cite{a8, a9}, where $r_g$ denotes the Schwarzschild radius. In \cite{b2} and \cite{b3}, researchers studied the effects of ISCO in the vicinity of Kerr BH and introduced these characteristics in GR.

Thorne and Novikov \cite{b4} determined the Kerr and Schwarzschild BH accretion discs efficiency. In \cite{b6} Johannsen created the accretion discs around such BHs, while Johannsen and Psaltis \cite{b5}  presented the Kerr-like metric. The geodesic structure and spherical orbits of charged particles near revolving, weakly magnetic BHs have been identified by Tursunov et al. \cite{b7}. Since the particles in the accretion disc revolve in stable orbits, oscillations in the radial, as well as vertical directions with epicyclic frequencies, arise if the particles are perturbed.

Because of this, understanding orbital and epicyclic frequencies is important for understanding the mechanisms of the accretion discs that surround BHs. Moreover, the accretion disc and geodesic structure have been analyzed in the literature for various BHs in \cite{b8}-\cite{ca7}.
The universal implications of non-linear electrodynamics (NED) theory, with the aim of examining the problem of universal evolution, as suggested by the Born-Infeld theory \cite{a63a}-\cite{a63c}. The study has emphasized the significance of NED in the field of cosmology, particularly regarding the time transition that both microscopic and macroscopic regions experience. In the last few years, there has been a significant amount of interest in cosmological models that incorporate NLED, as evidenced by the attention obtained in various studies \cite{a63d}-\cite{a63f}. The study of the NED phenomenon in celestial objects has experienced a notable expansion as a result of noteworthy findings \cite{a63g}-\cite{a63k}. The remarkable characteristics of Einstein's gravitational solutions and the NED field are revealed when examining their implications within the framework of the Big Bang cosmological model. The potential significance of NED fields in the cosmos cannot be understated. To have a comprehensive understanding of these solutions, it is essential to recognize the relationship between powerful NED fields. Previous research has investigated BHs characterized by many horizons within the framework of NED fields \cite{a63l}-\cite{a63p}. Recently, the nonlinear BH (RN-BH coupled with the NED parameter $\zeta$) solution has been computed in the framework of the NED field given in Ref. \cite{ac1}. Also, the first law of thermodynamics Smarr formula, and the physical characteristics of this BH are investigated in \cite{ac1}.

With the above motivations, this paper aims to investigate the properties of the circular geodesic and accretion disc surrounding RN-BH coupled with the NED parameter $\zeta$. To be conservative, we restrict our analysis to the polar coordinate system and the equatorial plane and circular orbits and calculate in detail the effects of the NED parameter $\zeta$ on  $r_{ph}$, $r_{mb}$, and $r_{isco}$. Moreover, the critical accretion is calculated using certain dynamic isothermal fluid parameters. The paper will be completed in the following manner. In Sec. II, we present a brief review of the RN-BH space-time coupled with the NED parameter $\zeta$. Sec. III is devoted to discussions on the general formulation for particle movement in the given subsections such as flux radiant energy, circular motion, oscillations, and stable circular orbits. In Sec. IV and its subsections, we determine the generic formulas for numerous dynamical parameters, critical flow speed, accretion for an isothermal fluid, and accretion rate. In Sec \textbf{V}, we examined the solution of the RN-BH coupled with the NED parameter $\zeta$ and a circular geodesic in the equatorial plane. In Sec \textbf{6} we discuss the summary of this article.

\section{Black hole spacetime in the NED model}


The action of the NED model is given by \cite{ac1}
\begin{equation}
S=\int d^{4}x\sqrt{-g}\left(\frac{R}{16\pi G}+\mathcal{L(F)}\right),\label{za1}
\end{equation}
where $G$ is the gravitational constant, $R$ indicates the Ricci scalar of the spacetime, and ${\cal L}(F)$ is the NED Lagrangian which is defined as
\bqn
{\cal L}({\cal F}) = -{\cal F} - \zeta \sqrt{-{\cal F}},
\eqn
with
\bqn
{\cal F} \equiv \frac{1}{4} { F}_{\mu\nu} { F}^{\mu\nu}
\eqn
being the Maxwell invariant and $\zeta$ the coupling constant. For later convenience, hereafter we set $G=1$. To obtain the spherical symmetric BH solution from Eq.~(\ref{za1}), one can start with the following metric ansatz,
\begin{equation}
ds^{2}=-f(r)dt^{2}+\frac{dr^{2}}{f(r)}+r^{2}d\Omega^{2},\label{za2}
\end{equation}
where
\bqn
d\Omega^{2}=d\theta^{2}+ \sin^{2}\theta d\varphi^{2},
\eqn
and $\psi(r)$ is the metric function.
The Einstein field equation can be written as
\begin{equation}
G^{\nu}_{\mu}=8\pi T^{\nu}_{\mu},
\end{equation}
where $G^{\nu}_{\mu}$ is the Einstein tensor and $T^{\nu}_{\mu}$ is the energy-momentum tensor of the nonlinear electromagnetic field which is of the form \cite{ac1}
\begin{equation}
T^{\nu}_{\mu}=\frac{1}{4\pi}(\mathcal{L} \delta^{\nu}_{\mu}-\mathcal{L_{F}}F_{\mu \gamma}F^{\nu \gamma}),
\end{equation}
where $\mathcal{L_{F}}= \frac{\partial\mathcal{L}}{\partial \mathcal{F}}$. The four components of Einstein's field equations are consistent and practically reduce to $G^{0}_{0}=8\pi T^{0}_{0}$ which can be explicitly written as
\begin{equation}
\frac{r f'(r)+f(r)-1}{r^{2}}=-\frac{q^{2}}{r^{4}}-\frac{4\zeta q\sqrt{q}}{3r}\ln (r),\label{za6}
\end{equation}
from (\ref{za6}) after some manipulation. By solving this equation one can obtain the following metric
\begin{equation}
f(r)=1-\frac{2M}{r}+\frac{q^{2}}{r^{2}}-\frac{4 q \sqrt{q}\zeta}{3r} \ln (r).\label{za7}
\end{equation}

\section{General formulation for the geodesic motion of test particles}

This section establishes the general formulation for the geodesic motion of the massive test particles by examining the RN-BH coupled NED parameter $\zeta$, which follows timelike geodesics. We assume $\xi_{t}=\partial_{t}$ and $\xi_{\phi}=\partial_{\phi}$ are killing vectors associated with fundamental constants, indicated as $E$ and $L$ (conserved energy and angular momentum) associated with the specified trajectory
\begin{equation}
E=-g_{\mu\nu} \xi_{t}^{\mu} u^{\nu}\equiv -u_{t},\label{a5}
\end{equation}
and
\begin{equation}
L=g_{\mu\nu} \xi_{\phi}^{\mu}u^{\nu}\equiv u_{\phi},\label{a6}
\end{equation}
where $u^{\mu}=\frac{dx^{\mu}}{d\tau}=(u^{t}, u^{r}, u^{\theta}, u^{\phi})$ is the four-velocity vector with $\tau$ being the affine parameter of the timelike geodesics. For timelike geodesics, the four-velocity vector $u^\mu$ fulfills the normalization condition $u^{\mu}u_{\mu}=1$ and then one obtains
\begin{equation}
[g_{rr}(u^{r})^{2}+g_{\theta\theta}(u^{\theta})^{2}]=[1-g^{tt}(u_t)^{2}-g^{\phi\phi}(u_{\phi})^{2}].\label{a7}
\end{equation}
From Eqs. (\ref{a5}), (\ref{a6}) and (\ref{a7}), in the equatorial plane (i.e. $\theta=\frac{\pi}{2}$), we have
\begin{equation}
u^t=-\frac{E}{f(r)},\label{a8}
\end{equation}
\begin{equation}
u^\theta=0,\label{a9}
\end{equation}
\begin{equation}
u^\phi=-\frac{L}{r^2},\label{b1}
\end{equation}
and
\begin{equation}
u^r=\sqrt{-f(r)\left(1-\frac{E^2}{f(r)}+\frac{L^2}{r^2}\right)}. \label{b3}
\end{equation}
Then Eq.~(\ref{b3}) gives
\begin{equation}
(u^r)^2+V_{eff}=E^2,\label{b4}
\end{equation}
where
\begin{equation}
V_{eff}=f(r)\left[1+\frac{L^2}{r^2}\right],\label{b5}
\end{equation}
is the effective potential. From the above result, it is obvious that the effective potential relies on the radial distribution, angular momentum, and space-time parameter $f(r)$.
The effective potential is incredibly helpful in the geodesic motion of the particles due to its ability to identify the position of ISCO by analyzing the local extrema of the effective potential.

\subsection{Circular Motion of Test Particles}

Let us first consider the circular motion of the massive particles in the equatorial plane $(\theta=\pi/2)$. For circular motion,  one has the conditions
\bqn
u^r=\frac{dr}{d\tau} =0\;\; {\rm and}\;\; \dot u^{r} = \frac{d^2r}{d\tau^2}=0.
\eqn
With these two conditions and by using Eq.~(\ref{b4}),  one has
\bqn
V_{eff}=E^2 \;\; {\rm and}\;\; \frac{d}{dr}V_{eff}=0.
\eqn
Then one can obtain the angular velocity $\Omega_{\phi}$,
specific energy $E$, and the specific angular momentum $L$ associated with the test particle that are provided respectively by
\begin{eqnarray}
\Omega_{\phi}^2=\frac{1}{2r}f'(r),\label{b6}
\end{eqnarray}
\begin{eqnarray}
E^2=\frac{2f^2(r)}{2f(r)-r f'(r)},\label{b7}
\end{eqnarray}
and
\begin{eqnarray}
L^2=\frac{r^3f'(r)}{2f(r)-r f'(r)}. \label{b8}
\end{eqnarray}
Frome Eqs.~(\ref{b7}) and (\ref{b8}), $E$ and $L$ should be real if
\begin{eqnarray}
2f(r)-r f'(r)>0. \label{c1}
\end{eqnarray}
From the above expression, the specific area of the circular orbit can be investigated. So, the inequality  (\ref{c1}), is essential for the existence of circular orbits. For bound orbits, one requires $E^2<1$ while the marginally bound orbit corresponds to $E^2=1$. Thus the marginally bound orbit satisfies
\begin{eqnarray}
2[f(r)-1]f(r)+r f'(r)=0.  \label{c2}
\end{eqnarray}
This equation determines the radius of the marginally bound orbit. On the other hand, it is easy to see that Eqs.~(\ref{b7}) and (\ref{b8}) diverge if
\begin{eqnarray}
-r f'(r)+2f(r)=0. \label{c3}
\end{eqnarray}
With this equation, one can determine the photon sphere radius which is essential for the investigation of gravitational lensing.

\subsection{Radiant Energy Flux and Circular Orbits}

The presence of stable circular orbits is dependent upon the local minima of the effective potential, which is attained when $\frac{d^{2} V_{eff}}{dr^{2}}>0$. From Eq. (\ref{b5}), we have
\begin{eqnarray}
\frac{d^2V_{eff}}{dr^2}=\left(1+\frac{L^2}{r^2}\right)f''(r)-\frac{4L^2}{r^3}f'(r)+\frac{6L^2}{r^4}f(r). \nb\\
\label{c4}
\end{eqnarray}
By employing the requirements $V_{eff} =0$, $\frac{dV_{eff}}{dr} =0$, and $\frac{d^{2}V_{eff}}{dr^{2}} =0$, one is able to calculate the radius of  ISCO, i.e.,$r_{isco}$.
Furthermore, the process of accretion is possible when $r<r_{isco}$. When particles fall from a state of rest to an infinite distance, accreting onto compact objects, they emit gravitational energy, which is converted into radiation. In \cite{h2}, the expression for the energy flux radiating across the accretion disc is formulated on the basis of the angular velocity $\Omega_{\phi}$, the specific energy $E$, and the specific angular momentum $L$ as
\begin{eqnarray}
K=-\frac{\dot{M} \Omega_{\phi},_{r}} {{4\pi\sqrt{-g} (E-L \Omega_{\phi})^2}}\int^{r}_{r_{isco}}(E-L \Omega_{\phi})L,_{r}dr, \nb\\
\label{c5}
\end{eqnarray}
where radiant flux is represented by $K$, mass accretion rate is $\dot{M}$, $\Omega_{\phi},_{r} \equiv \frac{d\Omega_{\phi}}{dr}$ and $g$ is the determinant of the metric tensor $g_{\mu\nu}$. $g$ is given by
\begin{eqnarray}
g=\det(g_{\mu\nu})=-r^4\sin^2\theta,  \label{c6}
\end{eqnarray}
In order to properly analyze our findings within the equatorial plane, we establish the relationship $\sin\theta=\sin\frac{\pi}{2}=1$. By applying Eqs. (\ref{b6}-\ref{b8}), we obtain
\begin{eqnarray}
K(r)&=&\frac{-\dot{M}}{4\pi r^4} \sqrt{\frac{r}{2f'(r)}} \nb\\
&&\times \frac{[r f'(r)-2f(r)][r f''(r)-f'(r)]}{[2f(r)+r f'(r)]^2}  \int^{r}_{r_{isco}} F(r)dr,  \nb\\
&& \label{c7}
\end{eqnarray}
where the function $F(r)$ is given by
\begin{eqnarray}
F(r)&=&\sqrt{\frac{r}{2f'(r)}} \Big[r f'(r)+2f(r)\Big]\nb\\
&& \times  \frac{r f(r)f''(r)-2r f'^2(r)+3f(r)f'(r)} {[2f(r)-r f'(r)]^2}.  \label{c8}
\end{eqnarray}
We assume that the accretion disc is in a state of thermal equilibrium, thus according to the Stefan-Boltzmann law, one has $K(r)=\sigma T^{4}(r)$ with $\sigma$ being the Stefan-Boltzmann constant, which is the relation between energy flux and temperature. Consequently, the radiation produced from the accretion disc is presumed to possess characteristics similar to those of black body radiation. By considering the thermal black body radiation, it is simple to determine the temperature distribution of the accretion disc by utilizing the given equation, from which one can figure out disc luminosity $L(\nu)$ where $\nu$ is the frequency of the radiated photons. The disc luminosity is determined by \cite{R3}
\begin{eqnarray}
L(\nu)=\frac{8 \cos\gamma}{\pi} \int^{r_{f}}_{r_{i}} \int^{2\pi}_{0} \frac{\nu_{e}^{3} r } {e^\frac{\nu_{e}} {T}-1} d\phi dr.\label{c9}
\end{eqnarray}
where $\gamma$ is the inclination angle of the accretion disc.
From the preceding result, we can see that the flux energy is denoted by $I(\nu)=L(\nu)/(4\pi)$. The maximum efficiency $\eta^{*}$ can be obtained by
\begin{eqnarray}
\eta^{*}=1-E_{isco}.\label{d1}
\end{eqnarray}
Here, $E_{isco}$ is the energy of the particles at ISCO. This proceeding relation holds when all emitted photons have the ability to escape infinity. When a fluid element experiences a perturbation, the resulting particles move, which corresponds to a circular orbit within the plane $\theta=\frac{\pi}{2}$.

\subsection{Oscillations}

In accretion processes, numerous types of oscillatory motion are seen as a result of restoring forces. The oscillatory motion in both the horizontal and vertical directions arises from the influence of restoring forces acting upon perturbations within the accretion discs. In an accretion disc, the number of restoring forces arises from the rotational motion of the disc in the presence of a vertical gravitational field.

When a fluid element moves radially, it returns to its state of equilibrium through the rotational motion of the fluid by virtue of the presence of a restoring force. It is worth noting that the gravitational force within accretion discs serves as a counterbalance to the centrifugal force due to central objects. The fluid element is dragged outside or inside and returned towards the original radius utilizing $\Omega_{r}$ epicyclic frequency, depending on whether the latter exceeds the former or vice versa. When the fluid element encounters a vertical perturbation within the plane $\frac{\pi}{2}$, the field of gravitation pinches the elements that are perturbed, causing it to return to its original equilibrium state. The element of fluid exhibits harmonic oscillations within the equatorial plane due to the presence of a restoring force, characterized by a vertical epicyclic frequency $\Omega_{\theta}$. Three different types of motion, harmonic vertical motion with a vertical frequency, circular motion with an orbital frequency, and radial motion with a radial frequency, are responsible for the behavior of particles within the accretion disc. Consequently, within the equatorial plane, we can examine the redial motion and vertical motion in the vicinity of circular orbits.

Now, let us investigate the radial and vertical motions that are represented by $\frac{1}{2}\left(\frac{dr}{dt}\right)^2=V_{eff}^{(r)}$ and $\frac{1}{2}(\frac{d\theta}{dt})^2=V_{eff}^{(\theta)}$, exhibited by the particles under consideration, where using Eq. (\ref{a7}) for explain the radial and vertical motions we consider $u^{r}=0$, $u^{\theta}=0$, respectively.  Considering $u^{r}=\frac{dr}{d \tau}=\frac{dr}{dt}u^{t}$ and $u^{\theta}=\frac{d\theta}{d\tau}=\frac{d\theta}{dt}u^{t}$, we can deduce the following equations,
\begin{eqnarray}
\frac{1}{2}\left(\frac{dr}{dt}\right)^{2}=-\frac{1}{2}\frac{f^3(r)}{E^2}\left[1-\frac{E^{2}}{f(r)}+\frac{L^{2}}{r^{2}\sin^{2}(\theta)}\right]=V_{eff}^{(r)} \nb\\\label{d2}
\end{eqnarray}
and
\begin{eqnarray}
\frac{1}{2}\left(\frac{d\theta}{dt}\right)^2=-\frac{1}{2}\frac{f^2(r)}{E^2 r^2}\left[1-\frac{E^2}{f(r)}+\frac{L^2}{r^2 \sin^2\theta}\right]=V_{eff}^{(\theta)}.\nb\\\label{d3}
\end{eqnarray}
In the equatorial plane, we shall investigate the vertical and radial epicyclic frequencies in the vicinity of circular orbits by considering small perturbations denoted as $\delta r$ and $\delta \theta$.  Differentiating Eq.~(\ref{d2}) and Eq.~(\ref{d3}) with respect to time $t$, one obtains
\begin{eqnarray}
\frac{d^2 r}{dt^2}=\frac{d V_{eff}^{(r)}}{dr}.\label{d3}
\end{eqnarray}
For a particle having a perturbation in its original radius at $r = r_{0}$, characterized by a deviation $\delta r = r - r_{0}$, the resulting equation can be expressed as follows
\begin{eqnarray}
\frac{d^2}{dt^2} (\delta r)=\frac{d^2 V_{eff}^{(r)}}{dr^2} (\delta r) \Rightarrow (\delta \ddot{r})+\Omega_{r}^2(\delta r)=0,\label{d4}
\end{eqnarray}
where $\Omega_{r}^2\equiv -\frac{d^2}{d r^2} V_{eff}^{(r)}$ and dots denotes derivatives with respect to $t$. Through an analogous methodology, when considering a perturbation in the vertical direction $\delta \theta=\theta-\theta_{0}$, we arrive at the following result
\begin{eqnarray}
\frac{d^2 (\delta\theta)}{dt^2}=\frac{{d^2 V_{eff}}^{(\theta)}}{dr^2} (\delta\theta) \Rightarrow (\delta \ddot{\theta})+\Omega_{\theta}^2(\delta \theta)=0,\label{d5}
\end{eqnarray}
where $\Omega_{\theta}^2\equiv -\frac{d^2}{d\theta^2} V_{eff}^{(\theta)}$. In the Equatorial plane, Eqs. (\ref{d2}) and (\ref{d3}) lead to
\bqn
\Omega_{r}^2&=&\frac{1}{2E^2 r^4} \Big\{\big[(r^2+L^2)3f(r)-2 E^2 r^2\big]r^2 f(r)f''(r) \nb\\
&&~~~~~~~~~~ +2r^2\big[(r^2+L^2)3f(r)-E^2 r^2\big]f'^2(r)\nb\\
&&~~~~~~~~~~  -6L^2 f^2(r)\big[2rf'(r)-f(r)\big]\Big\},\label{d6}
\eqn
and
\begin{eqnarray}
\Omega_{\theta}^2=\frac{f^2(r)L^2}{E^2 r^4}.\label{d7}
\end{eqnarray}
The prime notation in Eq. (\ref{d6}) denotes differentiation with respect to the radial coordinate $r$. The subsequent section presents a comprehensive investigation of the fundamental dynamical equations governing RN-BH coupled with NED.

\section{Basic Dynamical Equations}

In this section, we performed an analysis of the fundamental formalism for the accretion process around RN-BH  coupled with NED. For this purpose, we follow the fundamental formalism that was established by Babichev et al.  \cite{R4,R5}. Let us start by considering an ideal fluid that is characterized by its energy-momentum tensor
\begin{eqnarray}
T^{\mu\nu}=(\rho+p)u^{\nu}u^{\mu}+g^{\mu\nu}p. \label{d8}
\end{eqnarray}
The quantities denoted as $p$ and $\rho$ correspond to the pressure and energy density of the fluid, respectively. In the equatorial plane, the four-velocity $u^{\mu}$ can be described as follows
\begin{eqnarray}
u^{\mu}=\frac{dx^\mu}{d\tau}=(u^t,u^r,0,0),\label{d9}
\end{eqnarray}
where $\tau$ is the proper time. By combining the above equation with the normalization condition $(u^{\mu}u_{\mu}=1)$, we obtain
\begin{eqnarray}
u^t=\frac{\sqrt{f(r)+(u^r)^2}}{f(r)}.\label{e1}
\end{eqnarray}
The requirement $u^{t}>0$ indicates flow is forward, whereas for accretion (flow is inward), the assumption $u^{r}<0$ holds. In order to analyze the accretion process, it is necessary to calculate the conservation equation of energy-momentum and conservation equation of particle-number. The conservation of energy-momentum tensor reads $T_{;\mu}^{\mu \nu}=0$, which leads to
\begin{eqnarray}
T_{;\mu}^{\mu \nu}=\frac{1}{\sqrt{-g}}(\sqrt{-g}T^{\mu\nu})_{,\mu}+ \Gamma_{\alpha\mu}^\nu T^{\alpha\mu}=0,\label{e2}
\end{eqnarray}
where $\sqrt{-g}=r^{2} \sin\theta$, $\Gamma$ is Christoffel symbol's of 2nd kind and $(;)$ is the covariant derivative associated with the metric $g_{\mu\nu}$. By utilizing the BH metric, Eq. (\ref{e2}) can be transformed into the form of
\begin{eqnarray}
T_{,r}^{10}+\frac{1}{\sqrt{-g}} T^{10} (\sqrt{-g})_{,r}+2 \Gamma_{01}^{0}T^{10}=0. \label{e8}
\end{eqnarray}
By solving Eq.~(\ref{e8}), one has
\begin{eqnarray}
\frac{d}{dr}[(\rho+p)u^r r^2 {\sqrt{f(r)+(u^r)^2}}]=0. \label{e9}
\end{eqnarray}
By performing the integration on the above equation, we are able to derive the resulting expression
\begin{eqnarray}
(\rho+p)u^r r^2 {\sqrt{f(r)+(u^r)^2}}=C_{0}. \label{f1}
\end{eqnarray}
In Eq.~(\ref{f1}), $C_{0}$ represents the integration constant. Using the principle of conservation coupled with the four-velocity, as expressed by the equation $u_{\mu}T_{;\nu}^{\mu \nu}=0$, we attain
\begin{eqnarray}
&&(\rho+p) u_{;\nu}^{\mu} {u_\mu u^\nu}+(\rho+p)_{,\nu} u_{\mu} {u^\mu u^\nu} \nb\\
&&+(\rho+p) {u_{\mu} u^\mu} u_{;\nu}^\nu+p_{,\nu}g^{\mu\nu} u_\mu+p{u_\mu} g_{;\nu}^{\mu\nu}=0. \label{f2}
\end{eqnarray}
Since $g_{;\nu}^{\mu\nu}=0$ and utilizing $u^{\mu}u_{\mu}=1$, one gets
\begin{eqnarray}
(p+\rho){u_{;\nu}^\nu}+u^{\nu}{\rho,_\nu}=0,\label{f3}
\end{eqnarray}
and since $A_{;a}^b=\partial_a A^b+\Gamma_{ac}^b A^c$, we acquire
\begin{eqnarray}
u^{r} \rho_{,r}+[\Gamma_{0c}^0u^c+u_{,r}^r+\Gamma_{1c}^{1}u^c+\Gamma_{2c}^{2}u^c+\Gamma_{3c}^{3}u^c](\rho+p)=0. \nb\\
\label{f4}
\end{eqnarray}
From Eq. (\ref{f4}), we attain
\begin{eqnarray}
\frac{\rho'}{(\rho+p)}+\frac{u'}{u}+\frac{2}{r}=0,\label{fa5}
\end{eqnarray}
and solving it yields
\begin{eqnarray}
r^2 u^r \exp\left(\int{\frac{d\rho}{p+\rho}}\right)=-C_1,\label{f5}
\end{eqnarray}
where $C_{1}$ is the integration constant. Assuming $u^{r}<0$, it follows that the above expression also has a negative sign and we calculate
\begin{eqnarray}
(p+\rho)\sqrt{(u^{r})^2+f(r)}\exp\left(-\int{\frac{d\rho}{p+\rho}}\right)=C_2. \label{f6}
\end{eqnarray}
The constant of integration, indicated as $C_{2}$ and the flux mass equation is
\begin{eqnarray}
(\rho u^\mu)_{;\mu}\equiv \frac{1}{\sqrt{-g}}(\sqrt{-g} \rho u^{\mu})_{,\mu}=0. \label{f7}
\end{eqnarray}
Then utilizing Eq. (\ref{f7}), we get
\begin{eqnarray}
\frac{1}{\sqrt{-g}}(\sqrt{-g}\rho u^{\mu})_{,r}+\frac{1}{\sqrt{-g}}(\sqrt{-g}\rho u^{\theta})_{,\theta}=0. \label{f8}
\end{eqnarray}
Note that the term $\frac{1}{\sqrt{-g}}(\sqrt{-g}\rho u^{\theta})_{,\theta}$ in Eq. (\ref{f8}) can be omitted since we only focus on the equatorial plane. Therefore, the expression $\sqrt{-g}\rho u^{\mu}$ is considered to be constant, i.e.,
\begin{eqnarray}
\rho u^r r^2=C_3, \label{f9}
\end{eqnarray}
where $C_{3}$ represents the integration constant.

\subsection{Dynamical Parameters}

\begin{figure*}
\includegraphics[width=8.1cm]{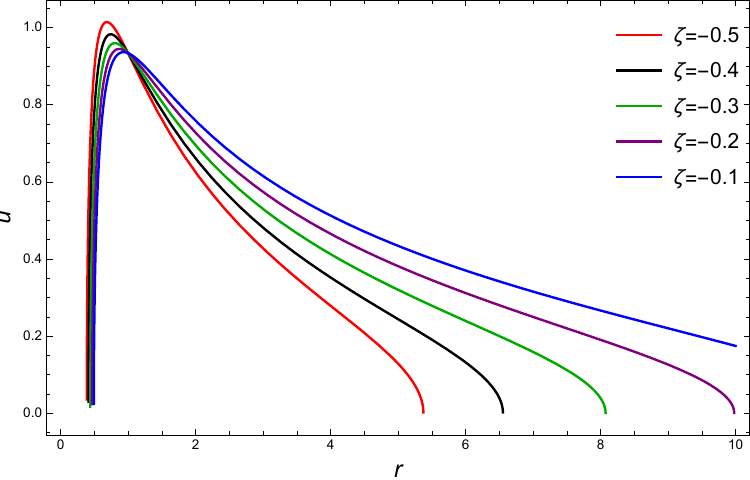}
\includegraphics[width=8.1cm]{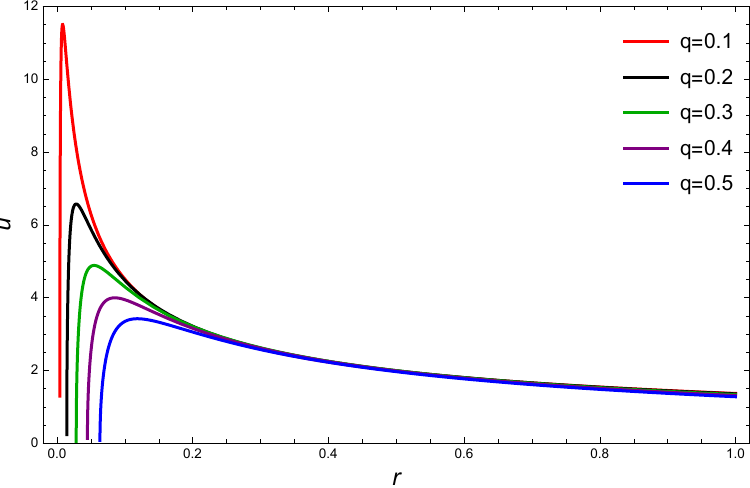}
\caption{The behaviors of $u$ as a function of $r$ for different values of electric charge $q$ and the NED paramater $\zeta$. The left panel corresponds to a fixed $q=1$ various values of $\zeta$ and the right panel corresponds to a fixed $\zeta=-1.5$ various values of electric charge $q$.} \label{fig: a}
\end{figure*}

Let us examine the accretion of the isothermal fluids characterized by the equation of state $p=k\rho$, where $k$ is the state function. For isothermal fluid $p\propto \rho$ it is necessary that the speed of sound remain constant in the accretion procedure. Moreover, from Eqs. (\ref{f5}), (\ref{f6}) and (\ref{f9}), one can determine
\begin{eqnarray}
\frac{\rho+p}{\rho} \sqrt{f(r)+(u^{r})^{2}}=C_4,\label{g1}
\end{eqnarray}
where $C_{4}$ represents the integration constant. Considering $p=k\rho$ and substitute it into Eq. (\ref{g1}), one has
\begin{eqnarray}
u=\frac{\sqrt{A_4^2-\frac{(k+1)^2 \left[3 \left(-2 M r+q^2+r^2\right)-4 \zeta  q^{3/2} r \ln r\right]}{3 r^2}}}{k+1} \label{g2}.
\end{eqnarray}
The graphs presented in \textbf{Fig.~\ref{fig: a}} illustrate the relationship between radial velocity $u$ and $r$. In the left panel of \textbf{Fig. \ref{fig: a}}, we analyze that initially, radial velocity increases to its maximum for a small BH radius with increasing the coupling parameter $\zeta$ and then moves toward a decreasing trend with increasing the BH radius. It is worthwhile to note that the maximum radial velocity attains at $\zeta=-0.5$. It is important to observe the impact of the coupling parameter, as radial velocity increases with increasing the coupling parameter for the given considered domain and show the stability of the system. In the right panel of \textbf{Fig. \ref{fig: a}}, we observe that the radial velocity attain maximum value at $q=0.1$ and decline gradually towards equilibrium position.

\begin{figure*}
\includegraphics[width=8.1 cm]{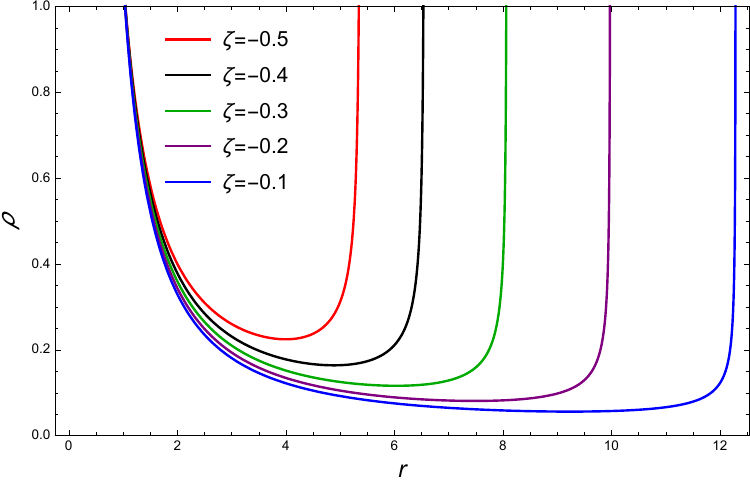}
\includegraphics[width=8.1 cm]{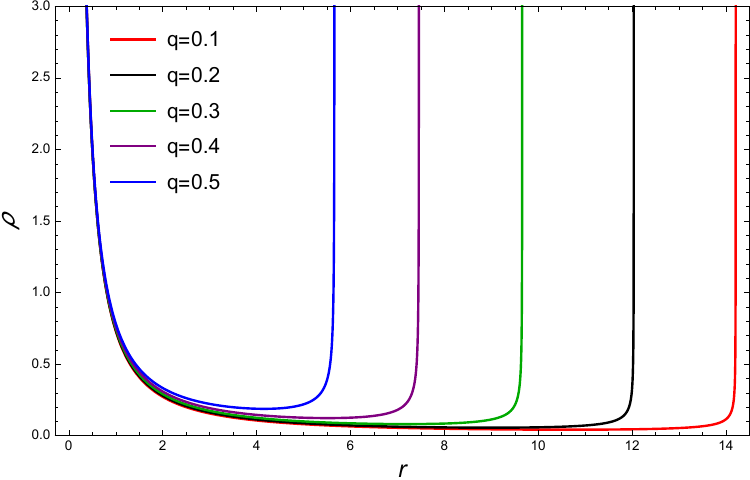}
\caption{ The behaviors of the fluid density $\rho$  as a function of $r$ for different values of electric charge $q$ and the NED paramater $\zeta$. The left panel corresponds to a fixed $q=1$ various values of $\zeta$ and the right panel corresponds to a fixed $\zeta=-1.5$ various values of electric charge $q$.} \label{fig: 2}
\end{figure*}

Now we can determine the density of fluid from Eq. (\ref{f9}), which is given by
\begin{eqnarray}
\rho=\frac{A_3 (k+1)}{r^2 \sqrt{A_4^2-\frac{(k+1)^2 \left(3 \left(-2 M r+q^2+r^2\right)-4 \zeta  q^{3/2} r \ln (r)\right)}{3 r^2}}}.\nb\\
\end{eqnarray}
The behavior of $\rho$ as a function of $r$ for different values of the electric charge and the NED parameter $\zeta$ are presented in Fig.~\ref{fig: 2}. From the left panel of \textbf{Fig. \ref{fig: 2}}, it is noted that initially fluid density declines for small BH radius and then grows rapidly toward maximum. Notably, the fluid density converges more rapidly for the smallest value of the BH parameter $ \zeta$. Also in the right panel of \textbf{Fig. \ref{fig: 2}}, observed opposite behavior as in \textbf{Fig. \ref {fig: 2}}. 

\subsection{Mass Evolution}

\begin{figure*}
\includegraphics[width=8.1 cm]{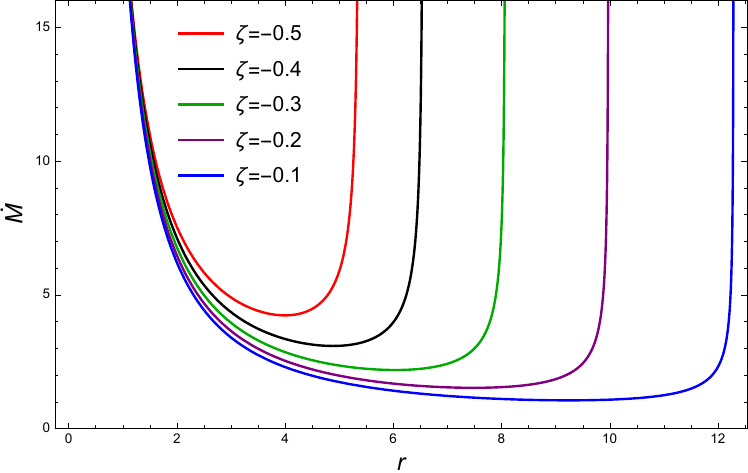}
\includegraphics[width=8.1 cm]{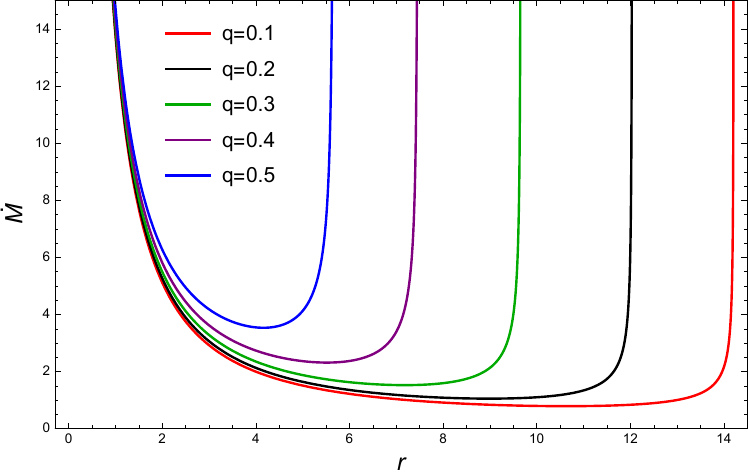}
\caption{The mass accretion rate $\dot{M}$ as a function of $r$ for different values of electric charge $q$ and the NED paramater $\zeta$. The left panel corresponds to a fixed $q=1$ various values of $\zeta$ and the right panel corresponds to a fixed $\zeta=-1.5$ various values of electric charge $q$.} \label{fig: 3}
\end{figure*}

Based on astronomical investigation, it is proposed that the mass of BH gradually varies over time due to various phenomena such as the emission of Hawking radiation and mass accreting around the BH. The mass accretion rate of RN-BH coupled with the NED parameter $ \zeta$ can be computed as $\dot{M}\equiv \frac{dM}{dt}=-\int T_{t}^{r}ds$ in which $ds=\sqrt{-g}d\theta d\phi$ and also $T_{t}^{r}=(p+\rho)u_{t}u^{r}$. As a consequence, accretion rate $\dot{M}$ is acquired by
\begin{eqnarray}
\dot{M}=-4\pi r^{2}u(p+\rho) \sqrt{u^{2}+f(r)} \equiv -4\pi C_{0}.\label{g4}
\end{eqnarray}
By asssuming $C_{0}=-C_{1}C_{2}$ and $C_{2}=(p_{\infty}+\rho_{\infty}) \sqrt{f(r_{\infty})}$, we have
\begin{eqnarray}
\dot{M}=4\pi C_{1}(p_{\infty}+\rho_{\infty}) \sqrt{f(r_{\infty})}M^{2}.\label{g5}
\end{eqnarray}
The evolution of time and BH mass can be obtained by considering $M_{i}$ is the initial mass and utilizing the Eq. (\ref{g5}), we obtained
\begin{eqnarray}
\frac{d M}{M^{2}}=\mathcal{F}t,\label{g6}
\end{eqnarray}
where $\mathcal{F}\equiv 4\pi C_{1}(p+\rho) \sqrt{f(r_{\infty})}$. From Eq.(\ref{g6}), we have
\begin{eqnarray}
M_{t}=\frac{M_{i}}{1-\mathcal{F} M_{i}t}\equiv \frac{M_{i}}{1-\frac{t}{t_{cr}}},\label{g7}
\end{eqnarray}
the expression for the time accretion $t_{cr}$ is calculated by using the formula $t_{cr}=[4\pi C_{1}(p+\rho)\sqrt{f(r_{\infty})}M_{i}]^{-1}$. According to Eq. (\ref{g7}), it is evident that at $t=t_{cr}$, the BH mass increases up to infinity within a finite time. So, the mass accretion rate of a BH is
\begin{eqnarray}
\dot{M}=4 \pi C_{1}(p+\rho)M^2.
\end{eqnarray}
\textbf{Figure \ref{fig: 3}} depicts the relationship between the mass accretion rate and the variable $r$. Under the influence of altered values of the parameter $\zeta$, it is observed that the accretion rate will initially decline along the BH radius $r$ and then increase to its maximum. Also, we note that the accretion rate $\dot{M}$ grow more rapidly for the smallest value of the BH parameter $\zeta$. Moreover, in the right panel of \textbf{Fig. \ref{fig: 3}}, we can see that the accretion rate gives the same behavior along BH radius $r$ as that in the left panel of \textbf{Fig. \ref{fig: 3}}, but the impact of BH charge $q$ is totally reversal to the case in the left panel.

\subsection{Critical Accretion}

The flow of fluid is static farthest from the BH, but it starts to move and accelerates inward due to the gravitational field exerted by the BH. When fluid flows inward, it reaches a sonic point where the velocity of the fluid is equivalent to the speed of sound. By utilizing Eqs.~(\ref{f9}) and (\ref{g1}), we have
\begin{eqnarray}
\frac{\rho'}{\rho}+\frac{u'}{u}+\frac{2}{r}=0,\label{g8}
\end{eqnarray}
and
\begin{eqnarray}
\frac{\rho'}{\rho}\left[\frac{d\ln(p+\rho)}{d\ln\rho}-1\right]+\frac{u u'}{u^{2}+f(r)}+\frac{1}{2} \frac{f'(r)}{u^{2}+f(r)}=0. \nb\\
\label{g9}
\end{eqnarray}
From Eq. (\ref{g9}), we attain
\begin{eqnarray}
\frac{d\ln u}{d\ln r}=\frac{D_{1}}{D_{2}},\label{h1}
\end{eqnarray}
where
\begin{eqnarray}
D_{1}=\frac{r f'(r)}{2(u^{2}+f(r))}-2V^{2},\label{h2}
\end{eqnarray}
and
\begin{eqnarray}
D_{2}=V^{2}-\frac{u^{2}}{u^{2}+f(r)}.\label{h3}
\end{eqnarray}
From Eqs. (\ref{h1})-(\ref{h3}), we attain
\begin{eqnarray}
V^{2}=\frac{d\ln(p+\rho)}{d\ln \rho}-1.\label{h4}
\end{eqnarray}
To determine the critical points, we assume  $D_{1}=D_{2}=0$ and we obtained
\begin{eqnarray}
V_{c}^{2}=\frac{r f'(r)}{4f(r)+r f'(r)},\label{h5}
\end{eqnarray}
and
\begin{eqnarray}
u_{c}^{2}=\frac{1}{4} r f'(r).\label{h6}
\end{eqnarray}
The index $c$ is referred to as the critical point. Note that the right-hand side of Eq. (\ref{h4}) is always positive. We compute the critical radius range by using the subsequent expression
\begin{eqnarray}
4f(r)+r f'(r)>0.\label{h7}
\end{eqnarray}
By utilizing Eq. (\ref{g2}), we acquire
\begin{eqnarray}
c_{s}^{2}=C_{4} \sqrt{[u^{2}+f(r)]^{-1}}-1.\label{h8}
\end{eqnarray}
The relation that represents the speed of sound is given by $c_{s}^{2}=\frac{d p}{d\rho}$.

\section{Circular equatorial geodesics}

\begin{figure*}
a) \includegraphics[width=8.1 cm]{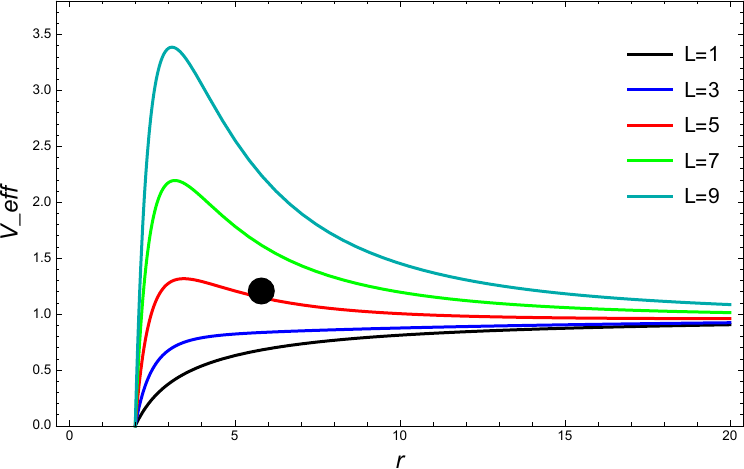}
b) \includegraphics[width=8.1 cm]{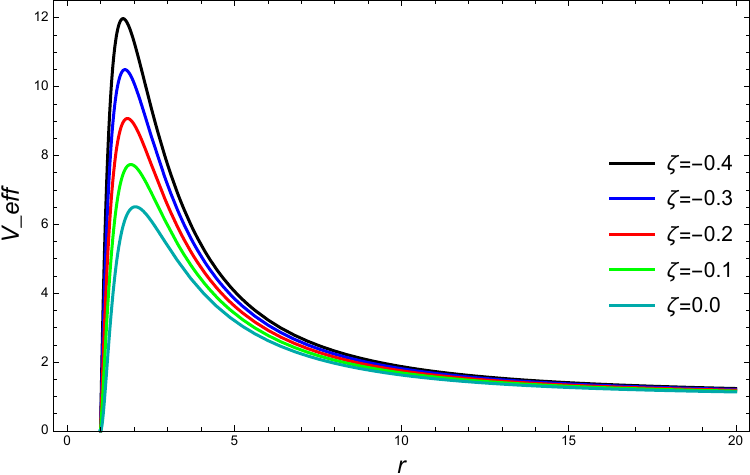}\\
c) \includegraphics[width=8.1 cm]{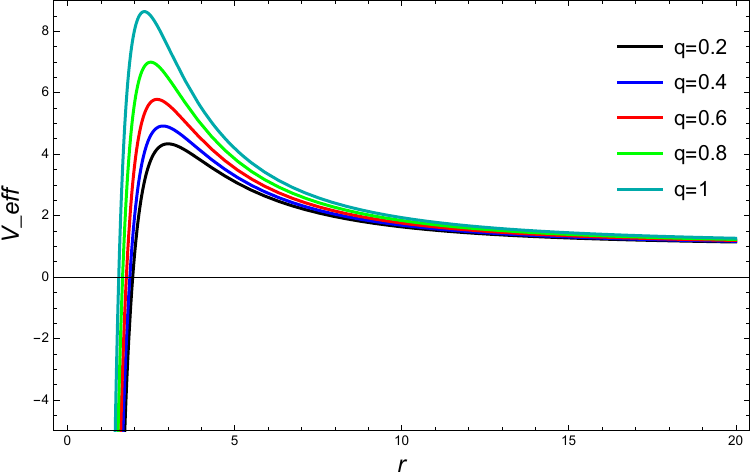}
\caption{ The illustration of $V_{eff}$ is a function of $r$.  (a) for $\zeta=-0.2$, $q=0.1$, and altered values of $L$ (b) for $q=1$, $L=10$ and various values of $\zeta$ c) for $\zeta=-1.5$, $L=10$ and distinct values of $q$.} \label{fig: 4}
\end{figure*}

The effective potential is necessary to examine circular geodesics in the plane $\theta = \frac{\pi}{2}$, which is obtained from Eq. (\ref{b5}) as
\begin{eqnarray}
V_{eff}=\left(1-\frac{2 M}{r}+\frac{q^2}{r^2}-\frac{4 \zeta  q \sqrt{q}\ln r}{3 r}\right)\left(1+\frac{L^2}{r^2}\right). \nb\\
\end{eqnarray}
The graphs presented in \textbf{Fig. \ref{fig: 4}} illustrate the behaviors of the effective potential as a function of the radial coordinate $r$ for different values of the specific angular momentum $L$, electric charge $q$, and the NED parameter $\zeta$. In \textbf{Fig. \ref{fig: 4} a}, one can observe the first extrema exist at $L = 5$ and no other extrema arise for $L<5$.  Additionally, the effective potential increases when angular momentum $L$ rises. The black dot within \textbf{Fig. \ref{fig: 4} a} gives the precise location of the ISCO, situated at $r=5.97347$. Furthermore, the effective potential $V_{eff}$ reveals two extrema for higher values of $L$. The stable and unstable circular orbits lie at the minimum and maximum of $V_{eff}$, respectively. From \textbf{Fig. \ref{fig: 4} b}, it is clear that as the value of BH parameter $\zeta$ rises, the effective potential decreases. In addition, from \textbf{Fig. \ref{fig: 4} c}, we observe that how the BH charge $q$ affects the effective potential $V_{eff}$ along $r$. In \textbf{Fig. \ref{fig: 4} c}, we can see $V_{eff}$ directly proportional to the charge $q$. We are currently interested in calculating ISCO for RN-BH coupled with NED because it forms the inner edges of the accretion disc, so the role of ISCO is significant. Unfortunately, we are unable to determine the ISCO analytically. Therefore, we will have to turn to numerical approaches by utilizing the general formula for ISCO provided in reference [1]. Whenever  $\zeta=-0.2$, $M=1$, and the charge $q=0.1$, the ISCO of BH is $r_{isco} =5.97347$. More detail regarding this can be found in \textbf{Table 1}.

In order to thoroughly investigate the accretion process, the ISCO is significant. It is also mandatory to conduct an analysis of other radii to obtain a comprehensive understanding. As previously mentioned, a circular orbit is present when the value of  $r>r_{ph}$. The particle's motion will demonstrate instability for small perturbations when the radial distance lies within the range of $r_{ph}<r<r_{isco}$. This relationship suggests that particles either escape to infinity or are dragged into the BH. If $r>r_{isco}$., the particle proceeds to move in stable circular orbits. Also, the photon sphere $r_{ph}$, circular orbit $r_{isco}$ and marginally bound orbit $r_{mb}$ are given in \textbf{Table 1.}

\begin{figure*}
(a)\includegraphics[width=8.1 cm]{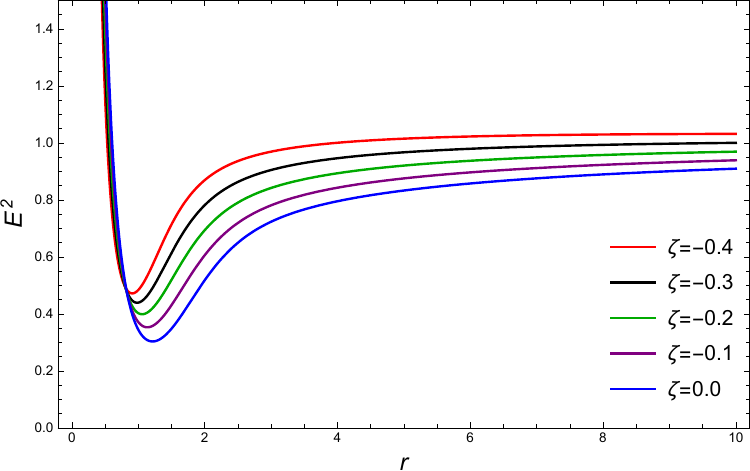}
(b)\includegraphics[width=8.1 cm]{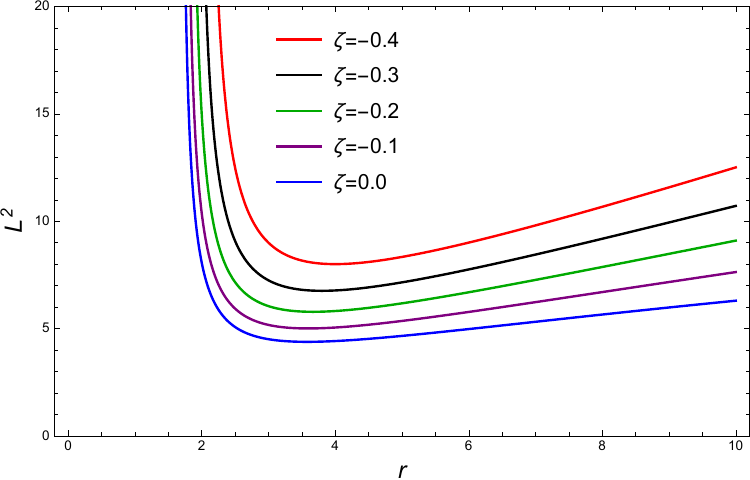}
(c)\includegraphics[width=8.1 cm]{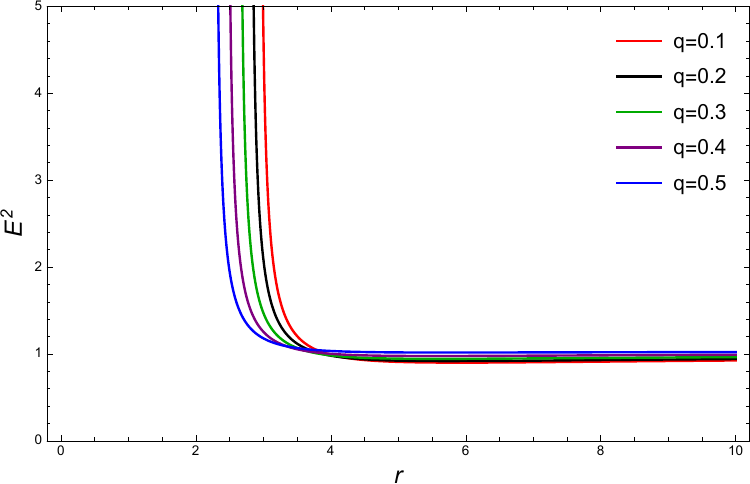}
(d)\includegraphics[width=8.1 cm]{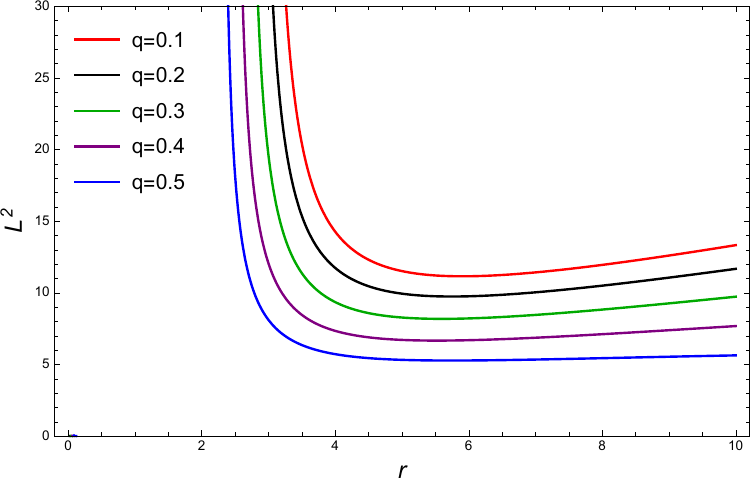}
\caption{ The profile of energy illustrated in left plots while in right plots angular momentum is depicted along $r$ for numerous values of BH parameters $\zeta$ and $q$.}. \label{fig: h}
\end{figure*}

In the plane $\theta=\frac{\pi}{2}$, one can derive the formulas for the following quantities
\begin{eqnarray}
E^{2}= \frac{1}{3r^2}\frac{\left[3 \left(r (r-2 M)+q^2\right)-4 \zeta  q^{3/2} r \ln r\right]^2}{3 r (r-3 M)+2 \zeta  q^{3/2} r-6 \zeta  q^{3/2} r \ln r+6 q^2}, \nb\\
\end{eqnarray}
\begin{eqnarray}
L^{2}=\frac{3 r^2 \left(3 M r-2 \zeta  q^{3/2} r+2 \zeta  q^{3/2} r \ln r-3 q^2\right)}{3 r (r-3 M)+2 \zeta  q^{3/2} r-6 \zeta  q^{3/2} r \ln r+6 q^2},
\end{eqnarray}
and
\begin{eqnarray}
\Omega_{\phi}^{2}=\frac{3 M r-2 \zeta  q^{3/2} r+2 \zeta  q^{3/2} r \ln r-3 q^2}{3 r^4}.
\end{eqnarray}
\textbf{Figure. \ref{fig: h}} depicts the specific energy and specific angular momentum profile along $r$. Also, the effects of BH parameters $\zeta$ and charge $q$ have been investigated in the given plots.
Presently, our focus to analyze the following quantities $E_{isco}$, $L_{isco}$, $\Omega_{isco}$, and $l_{isco}$ in ISCO. However, it is worth noting that an analytical identification of all these quantities is not possible. Therefore, the numerical calculation is outlined in the provided table.

\subsection{Radiant energy flux}

\begin{figure}
\includegraphics[width=8.1 cm]{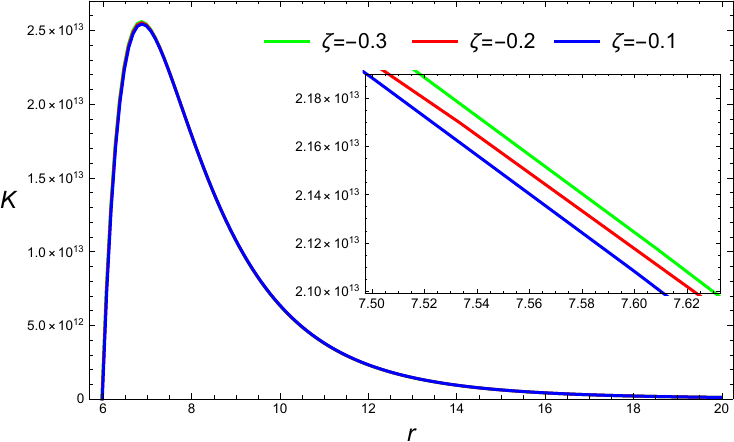}
\caption{The profile of energy flux $K$ along $r$, for different value of $\zeta$.} \label{fig: 5p}
\end{figure}

We begin to investigate the flux radiation emanating from the outermost layer of the disc in the plane $\theta=\frac{\pi}{2}$ by the utilization of the corresponding quantities $E$, $L$, and $\Omega_{\phi}$. The flux radiant energy associated with the accretion disc can be investigated through Eqs. (\ref{c7}) and (\ref{c8}), given as
\begin{widetext}
\bqn
K(r)&=&-\Bigg\{\sqrt{3} \dot{M} \sqrt{\frac{r^4}{3 M r-2 \zeta  q^{3/2} r+2 \zeta  q^{3/2} r \ln r-3 q^2}}\Big(9 M r-8 \zeta  q^{3/2} r+6 \zeta  q^{3/2} r \ln r-12 q^2\Big)\nb\\
&&~~~~~~~~~~~~~~~ \times \Big[3 r(-2 M r-M+r^2)+2 \zeta  q^{3/2} r-2 \zeta  q^{3/2} r (2 r+1) \ln r+3 q^2 (r+1)\Big]\Bigg\} \nb\\
&&~~ \times \Big[8 \pi  r^5(3 r (-2 M r+M+r^2)-2\zeta  q^{3/2} r+ 2 \zeta  q^{3/2} r (1-2 r) \ln r+3 q^2 (r-1))^2\Big]^{-1} \int^{r}_{r_{isco}}F(r)dr,
\eqn
where
\bqn
F(r)&=&\Bigg\{\sqrt{\frac{r^4}{9 M r-6 \zeta  q^{3/2} r+6 \zeta  q^{3/2} r \ln r-9 q^2}} \nb\\
&&~~~ \times  \Big[3 r(-2 M r+M+r^2)-2 \zeta  q^{3/2} r+2 \zeta  q^{3/2} r (1-2 r) \ln r+3 q^2 (r-1)\Big] \nb\\
&&~~~ \times \Big[2 \zeta  q^{3/2} r \ln (r)(3 r(r^2-4 M (r+2))+16 \zeta  q^{3/2} r-4 \zeta  q^{3/2} r (r+2) \ln r+3 q^2 (r+8))\\
&&~~~~~~~~~ -48 \zeta  q^{3/2} r (q^2-M r)+9(M q^2 r (r+8)+M r^2(r^2-2 M (r+2))-4 q^4)-16 \zeta ^2 q^3 r^2\Big]\Bigg\} \nb\\
&&~~~~ \times \Big(2 r^3(3 r (-2 M r-M+r^2)+2 \zeta  q^{3/2} r-2 \zeta  q^{3/2} r (2 r+1) \ln r+3 q^2 (r+1))^2\Big)^{-1}.
\eqn
\end{widetext}
In order to analyze the radiation flux behavior of the accretion disc surrounding BH for numerous values of the coupling parameter $\zeta$, as depicted in \textbf{Fig. \ref{fig: 5p}}, it is observed that as the parameter $\zeta$ grows, the flux energy of the accretion disc decreases.

\subsection{Radiant temperature}

\begin{figure}
\includegraphics[width=8.1cm]{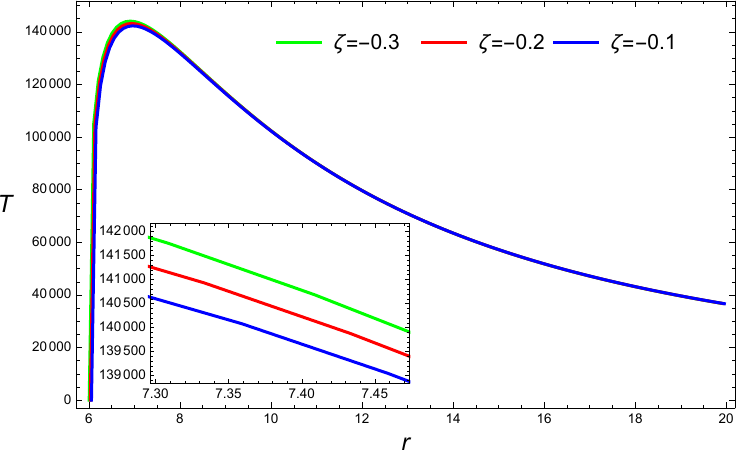}
\caption{The radiation temperature profile for various values of $\zeta$.}\label{fig: 6plus}
\end{figure}

It is speculated that the accretion disc is in thermal equilibrium, so the emission of radiation to the disc follows black body radiation principles. The Stefan-Boltzmann law, $K(r)=\sigma T^{4}$, established a relationship between energy flux and temperature. Here, $\sigma$ represents the Stefan-Boltzmann constant. The disc temperature is evaluated through the utilization of the BH parameter $\zeta$. In \textbf{Fig. \ref{fig: 6plus}}, we observe the temperature distribution on a disc for different values of $\zeta$ while keeping the parameter $q=0.1$ fixed. It is observed that the disc temperature drops as the parameter $\zeta$ assumes progressively larger values.

\subsection{Radiative efficiency}

The emission of radiation produced by the transformation of gravitational energy arises together with the gradual inward spiraling of the material comprising the disc towards its central region. The specific energy within the ISCO radius can be used to figure out the radiative efficiency, which is the ability of the central entity to convert mass at rest into radiation
\begin{eqnarray}
\eta=1- E_{isco}.
\end{eqnarray}
The numerical outcomes of ISCO, the photon sphere radius $r_{ph}$, the marginally bound orbit $r_{mb}$, the specific energy at ISCO denoted as $E^{2}_{isco}$, the specific angular momentum at ISCO indicated as $L^{2}_{isco}$, the specific angular velocity at ISCO referred to as $\Omega^{2}_{isco}$, the specific angular momentum at ISCO symbolized as $l^{2}_{isco}$, the maximum energy flux and the maximum temperature distribution are presented in \textbf{Table 1}.

\begin{table*}[ht]
    \caption{The numerical results are presented for $M = 1$, $q=0.1$ and various values of BH parameter $\zeta$.} 
    \centering 
    \begin{tabular}{c c c c c c c c c c c c} 
        \hline 
        $\zeta$ &ISCO & $r_{mb}$ \ &$r _{ph}$  \ &$ E^{2}_{isco}$ \ & $L^{2}_{isco}$ \ & $\Omega^{2}_{isco}$ \ & $l^{2}_{isco}$ & $K_{max}(r)$ & $T_{max}(r)$ &$\eta$\\ [0.7ex] 
        \hline 
              -0.3&5.967814&3.98041& 2.9789 &0.890526&11.7976&0.00467366 &13.248& $2.559\times10^{13}$ &   $1.439\times10^{5}$ &0.0563234\\ 
       -0.2  &5.9734741&3.98355&2.98369&0.889895&11.8536&0.00466815&13.3203&$2.551\times10^{13}$ &$1.43\times10^{5}$ &0.0566574\\
             -0.1  & 5.979193&3.98674&2.98849&0.889267&11.91&0.00466253&13.393 &         $2.54\times10^{13}$ &   $1.4215\times10^{5}$ &0.0569904\\
      \hline 
    \end{tabular}
    \label{table:nonlin}
\end{table*}

\subsection{Epicyclic frequencies}

\begin{figure*}
a)\includegraphics[width=8.1 cm]{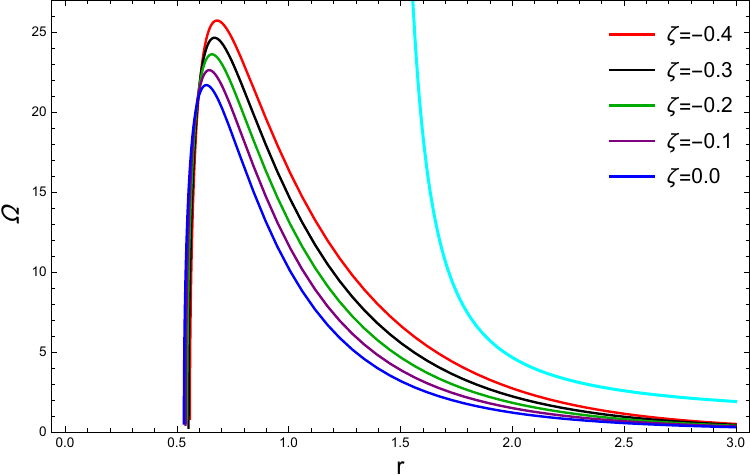}
b)\includegraphics[width=8.1 cm]{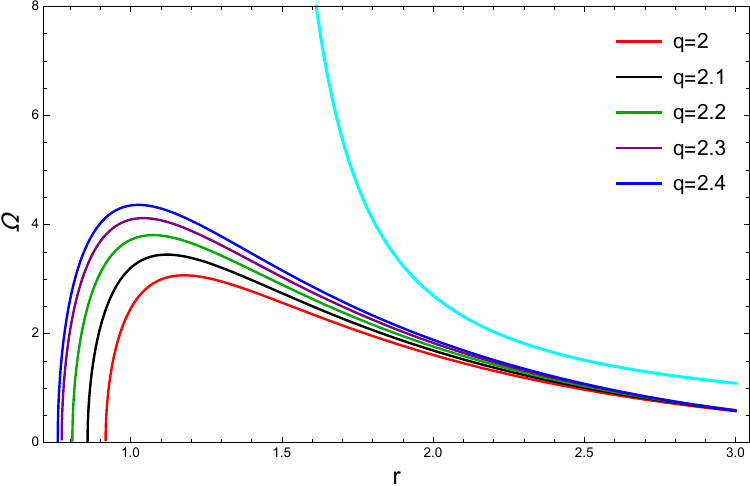}
\caption{The epicyclic frequencies depicted as a function of $r$ for altered values of BH parameter $\zeta$ in left plot and $q$ in right plot.} \label{fig: 6}
\end{figure*}

In the presence of perturbations within the equatorial plane, when the particles move along a circular orbit, that will produce small oscillations in the vertical and radial directions. The radial and vertical epicyclic frequencies are determined from Eqs. (\ref{d6}) and (\ref{d7}). Since the expressions of these quantities are a little bit lengthy, we are not going to present them explicitly here. Instead, we calculate them numerically and presented the results in In \textbf{Fig. \ref{fig: 6}}.
In \textbf{Fig. \ref{fig: 6}}, the profile of epicyclic frequencies $(\Omega_{\theta}, \Omega_{r})$ can be examined along the dimensionless radial coordinate $r$ for numerous values of coupling parameter $\zeta$ and charge $q$ of BH. In \textbf{Fig.~\ref{fig: 6}}, the cyan curve indicates the behavior of the vertical epicyclic frequency along radial coordinates $r$. It is worthwhile to note that the vertical frequency goes down as the radius $r$ goes up, while the remaining curves express patterns of radial frequency along with radial coordinate $r$. From the given figure, it is clear that the radial frequency increases to its maximum for small radii and then starts decreasing towards equilibrium for large radii. The radial frequency is maximum at $\zeta=-0.4$ in the given domain. Moreover, in \textbf{Fig.~\ref{fig: 6}}, it can be seen that vertical frequency has a similar behavior as discussed in \textbf{Fig.~\ref{fig: 6}}. While the radial frequency gives reversal behavior due to charge parameter $q$ as mentioned in \textbf{Fig.~\ref{fig: 6}}.

\section{Conclusions}

We investigate the process of accretion and particle geodesic motion surrounding the Reissner-Nordstr\"{o}m BH coupled with NED parameter $\zeta$, in the equatorial plane. The stability and circular geodesics of their orbits have been investigated, examining the oscillations that arise from perturbations, the existence of unstable orbits, and ultimately enabling the construction of a fundamental formulation for understanding accretion flow near the BH. Furthermore, the dynamical parameters, effective potential, typical radius, specific energy, epicyclic frequencies, specific angular momentum, emission rate, and mass accretion rate of the BH are determined. By establishing the state function $p = k\rho$ for isothermal fluid, one can deduce the general solutions within the framework of a Reissner-Nordstr\"{o}m BH coupled with (NED) parameter $\zeta$.

The transformation of the loci of unstable and stable circular orbits is the obvious outcome of the impact of the BH parameter $\zeta$ on the effective potential, as observed in our investigations. As the parameter $\zeta$ is enhanced, the $V_{eff}$ has associated decline, thereby allowing us to identify the precise location of the ISCO as depicted in \textbf{Fig. \ref{fig: 4} a}. The location of the radii, namely $r_{isco}$, $r_{ph}$, $r_{sin}$, and $r_{mb}$, within this particular space-time exhibits significant deviations from the Schwarzschild solutions. In \textbf{Table 1}, we presented an examination of the numerical outcomes associated with the ISCO, the radius of the photon sphere, the marginally bound, $E_{isco}$, $l_{isco}$, $\Omega_{isco}$ and the radioactive efficiency. In \textbf{Fig. \ref{fig: h}}, we observe how the parameter $\zeta$ and charge $q$ affected the angular momentum and energy of the BH.
Furthermore, we observe that as $\zeta$ increases, the efficacy of the accretion process increases. It is noticeable that as the BH parameter $\zeta$ grows the flux of radiation and the radiant temperature has associated decline, while the radiative efficiency increases.

We examine the characteristics of the fluid particle density, radial velocity, and accretion processes by employing the state parameter $k=0.5$ with considering isothermal fluid. It has been observed that the radial velocity attains maximum value at small radii for the parameters $\zeta$ and $q$ near the BH but farthest from the BH fluid has no radial velocity. The accretion process happens when fluid traverses from the critical points its speed matches with sound speed. So prior to the critical point, the flow of the fluid has a subsonic regime. So prior to the critical point, the flow of the fluid has a subsonic regime, but the flow becomes supersonic as it crosses that point near the BH due to a strong gravitational field. Upon examining the rate of accretion, we determined that its behavior extensively relies on fluid nature and BH parameters $\zeta$ and charge $q$. In the scenario of a normal fluid, the growth in mass accretion happens due to the immense gravitational field, and it has a maximum value near the BH. The rise in mass accretion occurs due to a positive deviation in the Schwarzschild BH scenario. Finally, circular orbits with their properties and epicyclic frequencies are examined in this paper. The vertical epicyclic frequency is a decreasing function of the radial distance $r$, without any extrema. The impact of BH parameters $\zeta$ and $q$ on radial frequency is considerable. We can see that the radial frequency attains its maximum value at a small BH radius $r$ as illustrated in \textbf{Fig. \ref{fig: 6}}. Also, we observe that radial frequency decreases due to enhancement in BH parameter $\zeta$ but as the value of charge $q$ grows, the radial frequency will also increase.

\section*{Acknowledgements}
The work of G. Abbas has been partially supported by the National Natural Science Foundation of China under project No. 11988101. He is grateful to the compact objects and diffused medium Research Group at NAOC led by Prof. JinLin Han for the excellent hospitality and friendly environment. He is also thankful to The Islamia University of Bahawalpur, Pakistan for the grant of study leave.
Tao Zhu is supported by the Zhejiang Provincial Natural Science Foundation of China under Grant No. LR21A050001 and LY20A050002,  the National Key Research and Development Program of China Grant No.2020YFC2201503, the National Natural Science Foundation of China under Grant No. 12275238 and No. 11675143, and the Fundamental Research Funds for the Provincial Universities of Zhejiang, China under Grants No. RF-A2019015.

\end{document}